# Unified Growth Theory Contradicted by the Economic Growth in Europe

Ron W Nielsen[1]

Environmental Futures Research Institute, Gold Coast Campus, Griffith University, Qld, 4222, Australia

December, 2015

Historical economic growth in Western and Eastern Europe is analysed. These regions should have produced the best and the most convincing confirmation of the Unified Growth Theory because they, and in particular Western Europe, were the centre of the Industrial Revolution, which according to Galor was the prime engine of economic growth. However, the data for Western and Eastern Europe show a remarkable disagreement with the Unified Growth Theory. There is no connection, whatever, between the data and the Unified Growth Theory. The data show that there was never a transition from stagnation to growth because there was no stagnation. Industrial Revolution, which should have the strongest influence in these regions, had absolutely no impact on changing the economic growth trajectories. The alleged remarkable or stunning escape from Malthusian trap did not happen because there was no trap. Unified Growth Theory does not explain the mechanism of the economic growth because its explanations are based on mythical features, which did not exist, the features contradicted by data. This theory needs to be either thoroughly revised or most likely replaced by a theory supported by a professional analysis of economic growth data.

**Introduction**

In the recent series of publications (Nielsen, 2015a, 2015b, 2015c) we have demonstrated that Unified Growth Theory (Galor, 2005a, 2011) is contradicted by the economic growth in Africa, Asia and in the countries of the former USSR. Earlier, we have demonstrated that this theory is contradicted by the world economic growth and by the economic growth in Western Europe (Nielsen, 2014). Implicitly, this theory is also contradicted by the mathematical analysis of the historical economic growth showing repeatedly that global, regional and national economic growths were hyperbolic (Nielsen, 2015d).

Any theory contradicted by just one set of good data should be either rejected or revised. Unified Growth Theory is contradicted repeatedly by data. However, what is even more remarkable is that this theory is *contradicted by the same data which were used during its development*. Such a paradox is probably rare in science but in this case it can be easily explained because the data were never properly analysed.

Data for the Gross Domestic Product per capita (GDP/cap), often used in the Unified Growth Theory, cause a significant problem with their interpretations. They are even more confusing

---

than the GDP data. In one of the publications we have explained how to understand the GDP/cap distributions and how they should be analysed (Nielsen, 2015e). These distributions cannot be simplified in the same way as the GDP distributions by using their reciprocal values but we have explained how these distributions can be also used to check the validity of the Unified Growth Theory.

Even though the GDP/cap distributions look complicated their interpretation is simple: they are just linearly-modulated hyperbolic distributions. Economic growth described by the GDP/cap data is slow over a long time and fast over a short time, in much the same way as the economic growth described by the GDP data but it is futile to try to determine the time when the growth changed from slow to fast. We have shown that the determination of this time is *impossible* because the growth described by the GDP/cap ratio increases monotonically. It is a lost effort to look for a time of a takeoff in the economic growth and any claim of the existence of such takeoffs is scientifically unjustified. It is based only on impressions but in science impressions have to be checked by a rigorous analysis of data.

There is no science without data but there is also no science without the correct analysis of data. Maddison published excellent set of data (Maddison, 2001, 2010), but they are repeatedly mutilated, frequently by displaying just four selected points and joining them by straight lines (Ashraf, 2009; Galor, 2005a, 2005b, 2007, 2008a, 2008b, 2008c, 2010, 2011, 2012a, 2012b, 2012c; Galor and Moav, 2002; Snowdon & Galor, 2008). Such a treatment of good data is unscientific and it can be hardly expected to lead to reliable interpretations. Hyperbolic distributions may be confusing but they are even more confusing if the data are not correctly displayed and analysed. The remarkable and convenient feature of hyperbolic distributions is that their analysis is simple if they are presented using their reciprocal values (Nielsen, 2014).

In the current publication we shall continue exploring the validity of Galor's fundamental postulate of the three regimes of growth: the Malthusian regime of stagnation, the post-Malthusian regime and the sustained-growth regime. According to Galor (2005a, 2008a, 2011, 2012a), Malthusian regime of stagnation commenced in 100,000 BC (the claim not supported by any data) and lasted until 1750 for developed countries. The post-Malthusian regime was allegedly between 1750 and 1870, and the sustained-growth regime from 1870. The alleged post-Malthusian regime overlaps the Industrial Revolution, 1760-1840 (Floud & McCloskey, 1994).

We shall now focus our attention of the economic growth in Western and Eastern Europe, the regions where the mechanisms of the economic growth discussed by Galor should be most prominently displayed. This is the region where economic growth should be in the excellent agreement with the Unified Growth Theory because in these regions, the Industrial Revolution, "the prime engine of economic growth" (Galor, 2005a, p. 212), should have been working exceptionally well and its effects should be prominently displayed in the economic growth data. If the numerous processes discussed in the Unified Growth Theory have any relevance in explaining the mechanism of the economic growth, their effects should be prominently manifested in Europe. We shall demonstrate that they are not. We shall demonstrate that the Unified Growth Theory is also remarkably strongly contradicted by the economic growth in Europe.

In our analysis we shall use the latest data published by Maddison (2010). Galor used the earlier publication (Maddison, 2001) but any of these publications can be used to show that the Unified Growth Theory is contradicted by Maddison's data.

As in our earlier publications, we shall use two ways of displaying data: (1) semilogarithmic display of the GDP data and (2) the display of the reciprocal values.

Again we shall remind that the hyperbolic growth is described by the simple mathematical formula:

$$S(t) = (a - kt)^{-1} \qquad (1)$$

where, in our case, $S(t)$ is the GDP while $a$ and $k$ are positive constants.

The reciprocal of a hyperbolic distribution is a straight line:

$$\frac{1}{S(t)} = a - kt \qquad (2)$$

If the reciprocal values of data follow a decreasing straight line, the growth is not stagnant but hyperbolic. However, it is not even necessary to show that the reciprocal values follow a decreasing straight line to demonstrate the lack of support for the concept of stagnation. To prove the existence of the epoch of stagnation it is necessary to prove the presence of random fluctuations often described as Malthusian oscillations. Such random fluctuations should be clearly seen not only in the direct display of data but also in the display of their reciprocal values. It they are absent then there is no support in the data for claiming the existence of the epoch of stagnation. However, if the reciprocal values of data follow a decreasing straight line, they show, or at least strongly suggest, that the growth was hyperbolic.

If the growth is slowed down, the reciprocal values of data are diverted to a less steep trajectory. If the straight line remains unchanged, then obviously there is no change in the mechanism of growth. It makes no sense to divide a straight line into two or three arbitrarily selected sections and claim different regimes of growth controlled by different mechanisms.

According to Galor, the transition from stagnation to growth occurred at the end of the Malthusian regime of stagnation and was marked by a "remarkable" or "stunning" escape from Malthusian trap (Galor, 2005a, pp. 177, 220), the escape he describes as a takeoff. For developed countries this alleged takeoff is supposed to coincide with the Industrial Revolution and in Europe it should be most clearly reflected in the data describing economic growth. We should see a prominent boosting in the economic growth, which should be detected in the semilogarithmic display of data but even more prominently in the display of their reciprocal values. We should see a clear and prominent *downward* change in the trajectory of the reciprocal values. We shall show that this prominent signature is missing in the data for the economic growth in Europe. We shall show that the data tell one story while the Unified Growth Theory tells another and diametrically opposite story.

**Analysis of data for Western Europe**

We shall analyse two sets of data for Western Europe: (1) the data for 12 countries and the data for the total of 30 countries. The 12 countries are made of Austria, Belgium, Denmark, Finland, France, Germany, Italy, the Netherlands, Norway, Sweden, Switzerland and the United Kingdom. According to Maddison (2010), in 2008, these 12 countries accounted for 85% of the total GDP of the 30 countries of Western Europe. The total of the 30 countries includes also Ireland, Greece, Portugal, Spain and 14 other small west European countries.

The reason for analysing these two groups separately is that the listed 12 countries represent the most advanced economies and consequently that for these countries we should expect the best agreement between the Unified Growth Theory and the data. The numerous stories presented in the Unified Growth Theory should be most prominently reflected and confirmed in these 12 countries.

Economic growth between AD 1 and 2008 in the 12 leading countries of Western Europe is shown in Figures 1 and 2. The growth in the total of 30 countries is shown in Figures 3 and 4.

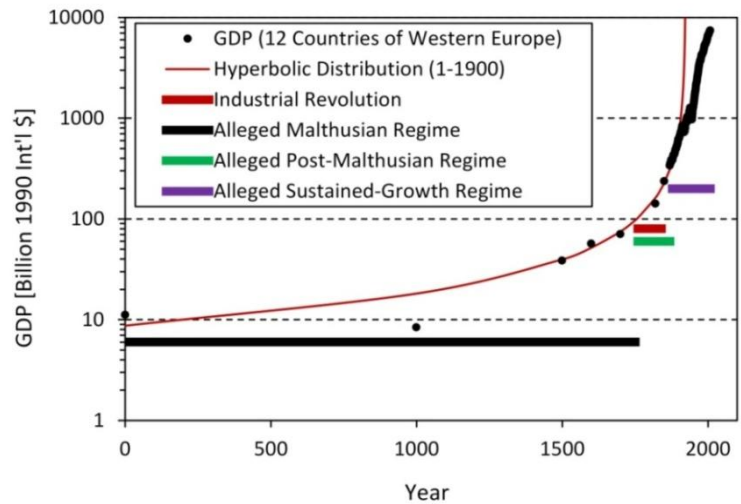

**Figure 1.** Economic growth in the 12 leading countries of Western Europe. The GDP is in billions of 1990 International Geary-Khamis dollars. There was no transition from stagnation to growth at any time. The growth was hyperbolic before and after the alleged transition. Industrial Revolution did not boost the economic growth. The "remarkable" or "stunning" escape from Malthusian trap (Galor, 2005a, pp. 177, 220) did not happen because there was no trap. Galor's three regimes of growth have no relevance to the description, let alone to the explanation, of the mechanism of the economic growth.

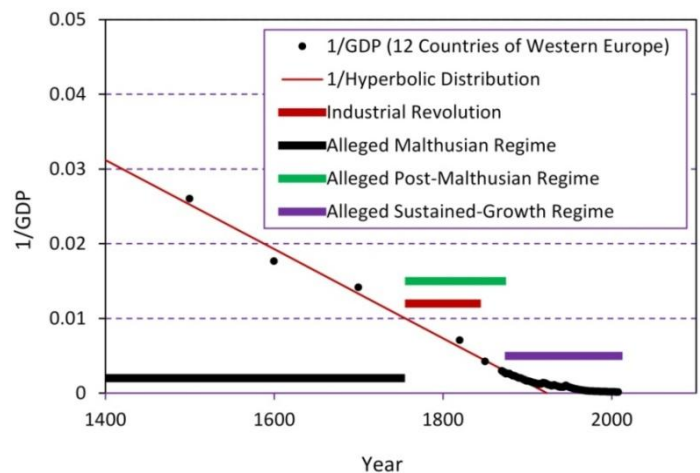

**Figure 2.** Reciprocal values of the GDP data, 1/GDP, for the economic growth in the 12 leading countries of Western Europe. Unified Growth Theory (Galor, 2005a, 2011) is contradicted by Maddison's data (Maddison, 2010). Galor's three regimes of growth have no relevance to the description, let alone to the explanation, of the mechanism of the economic growth. There was no transition from stagnation to growth at any time because there was no stagnation. There was no "remarkable" or "stunning" escape from the Malthusian trap (Galor, 2005a, pp. 177, 220) because there was no trap. Industrial Revolution did not boost the economic growth even in the countries where its effects should be most pronounced.

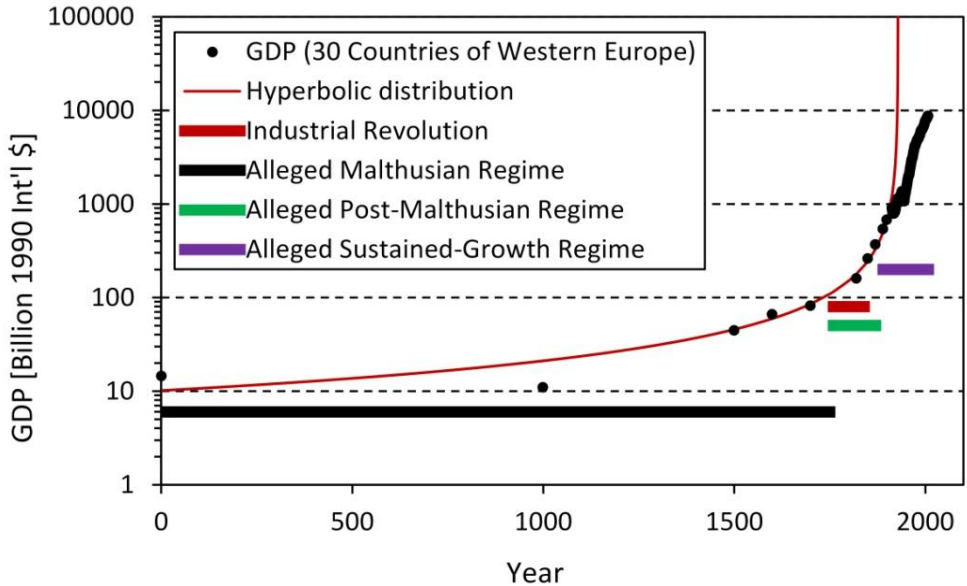

**Figure 3.** Economic growth in the total of 30 countries of Western Europe. The GDP is in billions of 1990 International Geary-Khamis dollars. The data give no support for the existence of the alleged Malthusian regime of stagnation. Industrial Revolution did not boost the economic growth in Western Europe. The "remarkable" or "stunning" escape from Malthusian trap (Galor, 2005a, pp. 177, 220) did not happen because there was no trap. Galor's three regimes of growth have no relevance to the description or to the explanation of the mechanism of the economic growth in Western Europe.

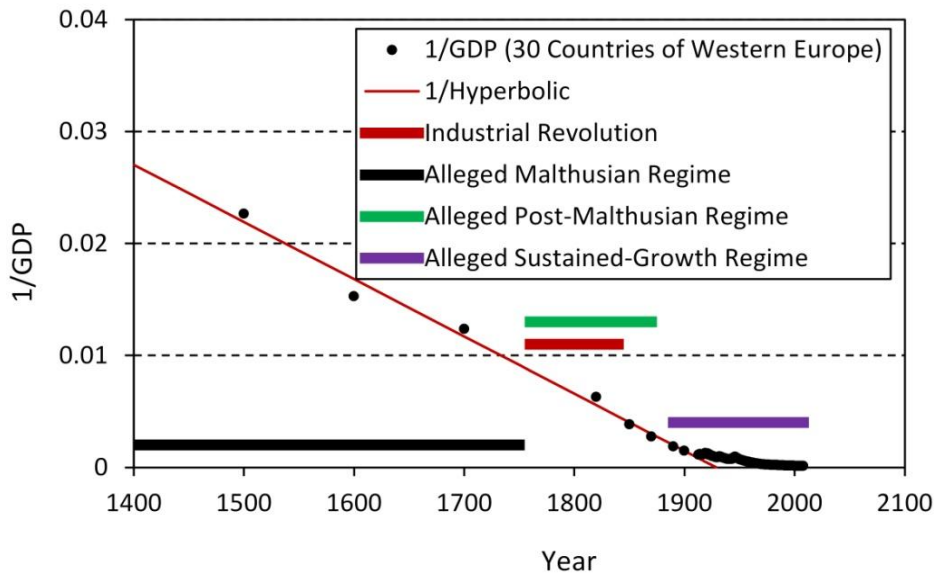

**Figure 4.** Reciprocal values of the GDP data, 1/GDP, for the economic growth in the total of 30 countries of Western Europe. Unified Growth Theory (Galor, 2005a, 2011) is contradicted by Maddison's data (Maddison, 2010). Galor's three regimes of growth have no expected connection with data. There was no transition from stagnation to growth at any time because there was no stagnation. There was no "remarkable" or "stunning" escape from the Malthusian trap (Galor, 2005a, pp. 177, 220) because there was no trap. Industrial Revolution did not boost the economic growth in Western Europe.

Hyperbolic parameters describing economic growth in the 12 countries of Western Europe are: $a = 1.147 \times 10^{-1}$ and $k = 5.961 \times 10^{-5}$. The corresponding singularity is in 1923 but the economic growth was diverted to a slower trajectory around 1900, bypassing the singularity by about 23 years.

Hyperbolic fit to the data is remarkably good between AD 1500 and 1900 and acceptable below AD 1500. The point at AD 1 is only 27% higher than the fitted distribution and the point at AD 1000 is 54% lower. The critical range of time for testing the Unified Growth Theory is from AD 1500 up. It is in this range of time that we should be able to see transition from stagnation to growth and later a transition to the alleged sustained growth regime.

The data presented in Figures 1 and 2 demonstrate clearly that there is no support for the existence of the alleged regime of Malthusian stagnation. However, there is a convincing support for the hyperbolic growth at least between AD 1500 and 1900, the range of time where the signature of Malthusian stagnation should be still clearly displayed for about 300 years. The data show that during that time economic growth was following a steadily-increasing hyperbolic trajectory. There is no sign of the existence of Malthusian stagnation.

Absolutely nothing had happened at the end of the alleged Malthusian regime. There was no transition from stagnation to growth at any time. On the contrary, around the beginning of the mythical regime of sustained-growth, when the economic growth was supposed to have been launched from stagnation to a fast-increasing trajectory, the growth started to be diverted to a slower trajectory.

Galor's Unified Growth Theory has no relevance to the description, let alone to the explanation of the mechanism of the economic growth, even in countries where his theory should be best fitted. Here, in the leading countries of Western Europe, where the effects of the Industrial Revolution should have been most prominently displayed in the economic growth data, where the "remarkable" and "stunning" escape from Malthusian trap (Galor, 2005a, pp. 177, 220) should be remarkably obvious, there are no signs of the impacts of the Industrial Revolution on the economic growth and no signs of any escape, remarkable or less-remarkable, because there was no trap. Economic growth was increasing undisturbed and unconstrained along a hyperbolic trajectory until around 1880 when it started to be diverted to a slower but still fast-increasing trajectory.

Stories presented so repeatedly by Galor in his theory have no relevance to explaining the mechanism of the economic growth even in these 12 leading countries of Western Europe. His three regimes of growth are dissociated from reality. His stories might be explaining or describing something else but they do not explain the mechanism of the economic growth.

Results of analysis of the economic growth in the total of 30 countries of Western Europe are presented in Figures 3 and 4. They lead to the same conclusions as for the 12 leading countries: Unified Growth Theory is contradicted by the economic growth data for Western Europe where the effects discussed by Galor should be most prominent. Unified Growth Theory presents stories, which are contradicted by the data describing economic growth.

Hyperbolic parameters describing economic growth in the total of 30 countries are: $a = 9.859 \times 10^{-2}$ and $k = 5.112 \times 10^{-5}$. The corresponding singularity is in 1929 but the economic growth was diverted to a slower trajectory around 1900, bypassing the singularity by about 29 years. The point at AD 1 is 42% higher than the calculated hyperbolic distribution and at AD 1000 it is 48% lover.

**Analysis of data for Eastern Europe**

Results of analysis of the economic growth in Eastern Europe, based on using Maddison's data (Maddison, 2010), are presented in Figures 5 and 6. Hyperbolic parameters fitting the data are: $a = 7.749 \times 10^{-1}$ and $k = 4.048 \times 10^{-4}$. The point at AD 1 is 51% higher than the calculated curve. The singularity is in 1915 but the economic growth was diverted to a slower trajectory around 1890, bypassing the singularity by 25 years.

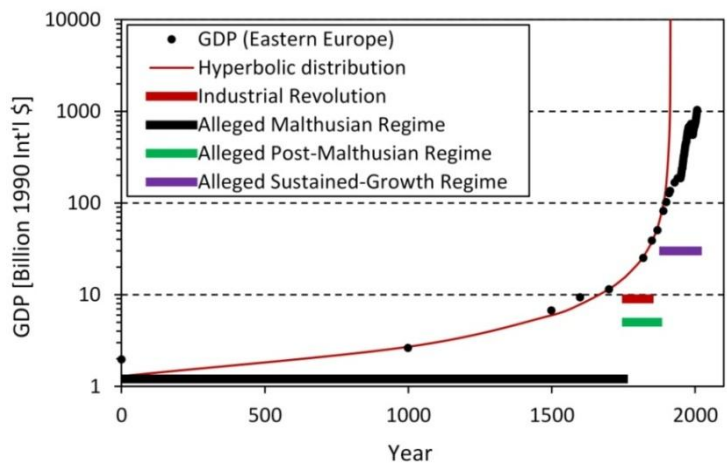

**Figure 5.** Economic growth in Eastern Europe. The GDP is in billions of 1990 International Geary-Khamis dollars. Galor's three regimes of growth have no relevance to the description, let alone to the explanation, of the mechanism of the economic growth. Unified Growth Theory is contradicted by data. The alleged Malthusian regime of stagnation did not exist. Industrial Revolution did not boost the economic growth in Eastern Europe. The "remarkable" or "stunning" escape from Malthusian trap (Galor, 2005a, pp. 177, 220) did not happen because there was no trap.

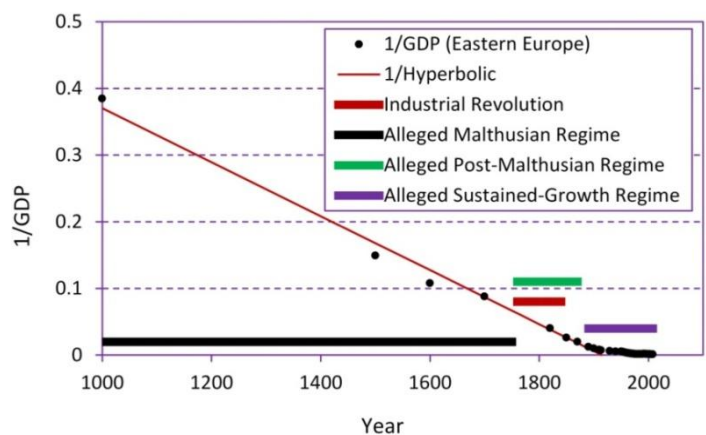

**Figure 6.** Reciprocal values of the GDP data, 1/GDP, for the economic growth in Eestern Europe. Unified Growth Theory (Galor, 2005a, 2011) is contradicted by Maddison's data (Maddison, 2010). Galor's three regimes of growth have no expected connection with data. There was no transition from stagnation to growth at any time because there was no stagnation. There was no "remarkable" or "stunning" escape from the Malthusian trap (Galor, 2005a, pp. 177, 220) because there was no trap. Industrial Revolution did not boost the economic growth in Eastern Europe. Galor's theory has no relevance to the description, let alone to the explanation, of the mechanism of the economic growth.

Unified Growth Theory is clearly contradicted by the economic growth data for Eastern Europe. The epoch of Malthusian stagnation did not exist. There was no transition from stagnation to growth at any time because the growth was hyperbolic. There was no "remarkable" or "stunning" escape from the Malthusian trap (Galor, 2005a, pp. 177, 220) because there was no trap. Industrial Revolution did not boost the economic growth in Eastern Europe.

There was also no boosting of the economic growth at the time of the transition from the alleged post-Malthusian regime to the alleged sustained growth regime. Soon after the commencement of this phantom sustained growth regime, economic growth in Eastern Europe started to be diverted to a slower trajectory. Galor's regimes of growth are clearly dissociated from data.

**Conclusions**

It is remarkable how Galor's Unified Growth Theory is dissociated from reality and how it is repeatedly contradicted by the analysis (Nielsen, 2014, 2015a, 2015b, 2015c, 2015d, 2015e) of Maddison's data (Maddison, 2001, 2010).

Western and Eastern Europe would be expected to give the strongest support for the Unified Growth Theory, because these regions should be expected to have been most strongly affected by the Industrial Revolution, the alleged "prime engine of economic growth" (Galor, 2005a, p. 212). However, it is remarkable how this theory is dissociated from data even for these regions. The contrast is amazing.

The theory tells one set of stories; the data present a diametrically different account. The theory claims a prolonged epoch of stagnation; the data show a steadily-increasing hyperbolic growth. The theory claims the existence of Malthusian trap; the data show that there was no trap. The theory claims a spectacular escape from the Malthusian trap; the data show an undisturbed hyperbolic growth. The theory claims a dramatic transition from stagnation to growth; the data show that there was no transition because there was no stagnation. The theory claims that the Industrial Revolution was the primary engine of economic growth; the data show that this engine had absolutely no effect on the economic growth trajectories. The theory claims a transition to a sustained growth regime; the data show that the economic growth was always sustained.

Concepts presented by the Unified Growth Theory are repeatedly and consistently contradicted by data. The described mechanism of growth has no reflection in data. The theory present stories, which are not supported by data.

This theory contains also dangerously misleading information. It claims that after the ages-long epoch of stagnation we have now entered the sustained-growth regime. It presents an image of a final and long-awaited liberation from Malthusian trap and of a transition to the unconstrained and sustained economic growth promising the ever-increasing prosperity. The data present the opposite picture. Not only do they show that the past economic growth was sustained but also that the current economic growth, even though diverted to slower trajectories, continues to increase too fast, pointing to the insecure future (Nielsen, 2015f), which can be also detected in the discussed here economic growth in Western and Eastern Europe.

Unified Growth Theory needs to be thoroughly revised but most likely replaced by a new theory to bring it in agreement with data. Data have to be properly analysed but the data describing hyperbolic growth have to be treated with extra care because they can be strongly misleading. Quoting certain carefully selected and isolated numbers from such distributions, as it is repeatedly done in the Unified Growth Theory, or plotting them in a strongly suggestive manner (Ashraf, 2009; Galor, 2005a, 2005b, 2007, 2008a, 2008b, 2008c, 2010, 2011, 2012a, 2012b, 2012c; Galor and Moav, 2002; Snowdon & Galor, 2008) can be hardly expected to lead to reliable conclusions. Unified Growth Theory is strongly, if not entirely, based on such unscientific use of data. It is, therefore, not at all surprising that the data, when rigorously analysed, are found to be in such a clear contradiction of this theory.

Unified Growth Theory needs to be thoroughly revised. Most likely, however, a new theory will have be to be proposed, a theory based on a scientific analysis of empirical evidence and in particular on a scientific analysis of Maddison's data (Maddison, 2001, 2010).